\documentstyle[prl,aps,epsfig,twocolumn]{revtex}
\begin{document} 
\draft \flushbottom \twocolumn[
\hsize\textwidth\columnwidth\hsize\csname @twocolumnfalse\endcsname

\title{Insulating, superconducting and large-compressibility phases 
in nanotube ropes}
\author{J. V. Alvarez and J. Gonz\'alez   \\}
\address{
        Instituto de Estructura de la Materia.
        Consejo Superior de Investigaciones Cient{\'\i}ficas.
        Serrano 123, 28006 Madrid. Spain.}

\date{\today}
\maketitle
\begin{abstract}
\widetext
The superconducting properties of carbon nanotube ropes are 
studied using a new computational framework  
that incorporates the renormalization of intratube interactions
and the effect of intertube Coulomb screening. This method allows
to study both the limits of thin and thick ropes ranging from 
purely one-dimensional physics to the setting of three-dimensional 
Cooper-pair
coherence, providing good estimates of the critical temperature
as a function of the rope physical parameters. We discuss 
the connection of our results with recent experiments.

\end{abstract}

]

\narrowtext 
\tightenlines


{\bf Introduction.}--- Carbon nanotubes offer nowadays 
a great potential for
technological applications. They display semiconducting or metallic
properties depending on the helicity of the tubule\cite{saito}, 
what is a useful feature to build electronic devices.
The electron correlations are significant in the metallic
nanotubes\cite{bal,eg,kane,yo}, explaining the experimental 
observation of features characteristic of Luttinger liquid
behavior\cite{exp,yao}. 
Quite remarkably, superconducting correlations have been also 
observed, in the form of the proximity effect\cite{kas,marcus} as 
well as in the sharp decrease of the resistance at low 
temperatures\cite{sup}.

The superconducting transitions reported in Ref.
\onlinecite{sup} have been measured in various samples of
nanotube ropes. These are supposed to be made of a large number
($\sim 300$) of single-walled nanotubes, and in all cases the 
critical temperature has been found below 1 K. In a different
kind of experiment\cite{chi}, strong superconducting correlations 
have been observed in individual nanotubes of very small diameter 
placed in a zeolite matrix. It has been claimed that such 
correlations should correspond to a critical temperature of 
the order of $\sim 15 \; {\rm K}$\cite{chi}. 

An explanation of the superconductivity (SC) in the 
nanotube ropes has been already proposed \cite{th1,th2}.
It has been argued that,
as long as the Cooper pairs are formed at zero total momentum,
they do not find the obstacle that single electrons have
to tunnel between neighboring nanotubes due to the misalignement
of the lattices\cite{mkm,avour}. The existence of a superconducting 
transition is possible in the ropes as the coherence in the 
transverse directions is established through the tunneling of 
Cooper pairs between the nanotubes\cite{th2}. 
In this scenario it has been assumed that the screening of the Coulomb
repulsion is the dominant effect in large nanotube ropes and 
that the backscattering and Umklapp interactions are   
not essential to explain the SC of these systems.

In this Letter we propose a new  computational 
framework to deal simultaneously with the Coulomb screening due 
to the intertube interactions and the scaling of all the intratube 
interactions in the low-energy theory. In this way we will 
obtain a precise estimate of the critical temperatures that can
be reached. 
The interplay between all the possible interactions in the 
scaling procedure will allow us to investigate the development
of 3D coherence in the rope, as well as to discover new phases 
which correspond to different charge instabilities in the system.

\begin{figure}
\begin{center}
\mbox{\epsfxsize 4.2cm \epsfbox{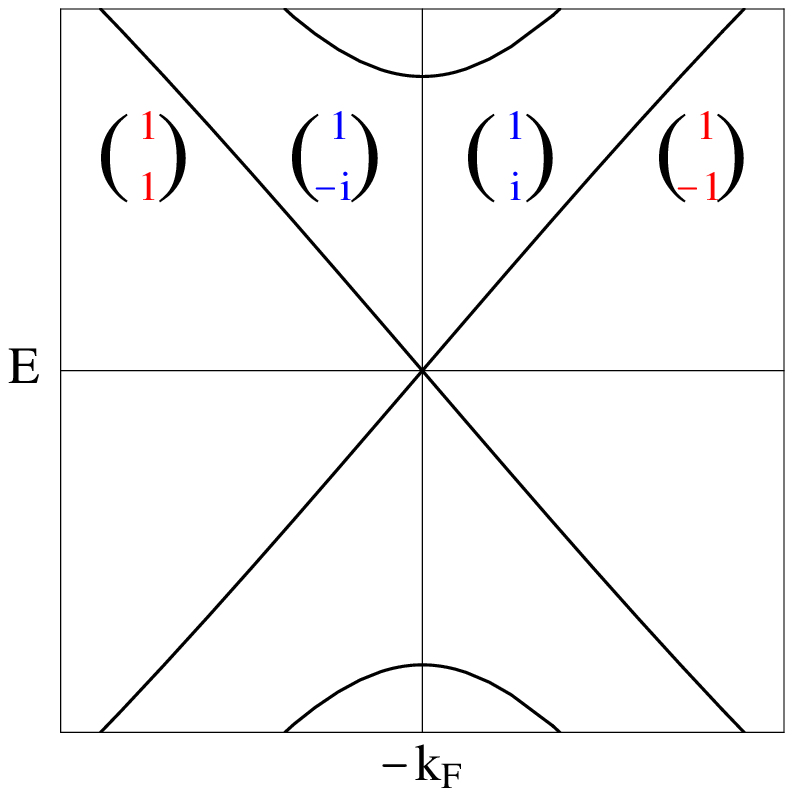} 
\epsfxsize 4.2cm \epsfbox{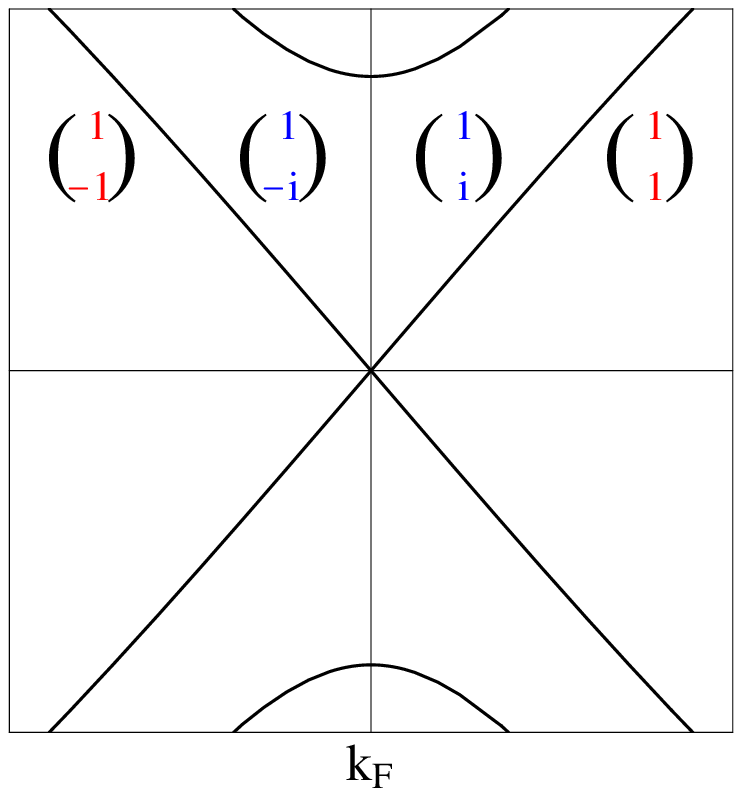}}
\end{center}
\caption{Plot of the low-energy linear branches of carbon nanotubes.
The spinors in the inside (outside) correspond to 
the relative electron amplitudes in the two sublattices of
graphene rolled up to form
a zigzag (armchair) nanotube.}
\label{bran}
\end{figure} 
 
{\bf Electron-phonon interactions.}--- The effective 
interaction coming from the exchange of phonons 
may lead to attraction or repulsion between the electrons,
depending on the particular
channel of interaction considered\cite{jdd,cara}.
We will follow the convention of classifying the four-fermion
interactions into different channels with respective coupling
constants
$g_i^{(j)}$\cite{berk}. The lower index discerns whether the
interacting particles shift from one Fermi point to the other
$(i=1)$, remain at different Fermi points $(i=2)$, or they
interact near the same Fermi point $(i=4)$. The upper label
follows the same rule to classify the different combinations of
left-movers and right-movers, including the possibility of
having Umklapp processes $(j=3)$.

Let us focus on the opposite cases of armchair and metallic
zigzag nanotubes\cite{saito}. All of them
have two pairs of linear branches in their band structure,
crossing at respective points at the Fermi level (for undoped
nanotubes) as shown in Fig. \ref{bran}.
Obviating the momentum dependence
of the electron modes, the relative amplitudes between points in 
the two sublattices of the honeycomb lattice are given by the
spinors in Fig. \ref{bran} \cite{dirac}. 


Using the fact that the electron modes with
opposite momenta have complex conjugate amplitudes, it can be
shown that the contributions from phonon exchange to  
$g_1^{(1)}, g_1^{(2)}, g_2^{(1)}, g_2^{(2)}$ and 
$g_4^{(1)}$ are negative, for frequencies below the 
characteristic phonon energy. However, the contributions from
Umklapp processes to $g_1^{(3)}, g_2^{(3)}$ and $g_4^{(3)}$ have
the opposite sign, since the respective products of 
electron-phonon couplings turn out to be negative\cite{cara}.
We recall that the electron-phonon couplings are given by a sum
involving products of electron modes in nearest-neighbor
sites\cite{jdd}.
Thus, the contributions to $g_4^{(2)}$ and $g_2^{(4)}$ have 
attractive character in the zigzag nanotubes, due to the 
relation of conjugation between the electron modes, but they 
have repulsive character in the armchair nanotubes, since only
one of the two scattered
modes reverses its sign there when shifting to a
neighboring site.

The signs of the different contributions are based on symmetry 
rules, and they hold irrespective of whether
the exchanged phonons are optical or acoustic. The energy of the
latter at momentum $2k_F $ is close to the energy of optical
phonons near $k = 0$, so that we will assume equal strength in
absolute value, $|g|$, for all the four-fermion interactions arising
from the exchange of phonons with a typical Debye frequency
between 0.1 and 0.2 eV.

We will deal with effective interactions coming
from the exchange of phonons within the same nanotube. Interactions 
mediated by the exchange of phonons in different nanotubes require 
mainly the coupling of out-of-plane phonons, which have
a much lower energy scale. These effects have been studied in Ref.
\onlinecite{dme}, with the result that the intratube superconducting
order parameter is more relevant than the intertube one, when the
system is in the regime with dominant attractive interactions.

{\bf Coulomb screening.}--- Regarding the Coulomb 
interaction, it has been shown that the
strength of the backscattering and Umklapp processes mediated by
the Coulomb potential is reduced by a relative factor
$\sim 0.1 \; a/R$, in terms of the ratio of the lattice spacing $a$
to the nanotube radius $R$\cite{bal,eg,kane}.
This means that, for typical nanotubes in a rope, these processes
can be safely neglected in favor of the contributions from phonon
exchange, whenever the strength in absolute value $|g|$ 
of the latter is above $\sim 0.05 v_F$.
Therefore, the competition between the long-range Coulomb
interaction and the effective interaction from phonon exchange
takes place in the channels with small momentum transfer
corresponding to the couplings $g_2^{(2)}, g_2^{(4)}, g_4^{(2)}$
and $g_4^{(4)}$.

\begin{figure}
\begin{center}
\mbox{\epsfxsize 8.5cm \epsfbox{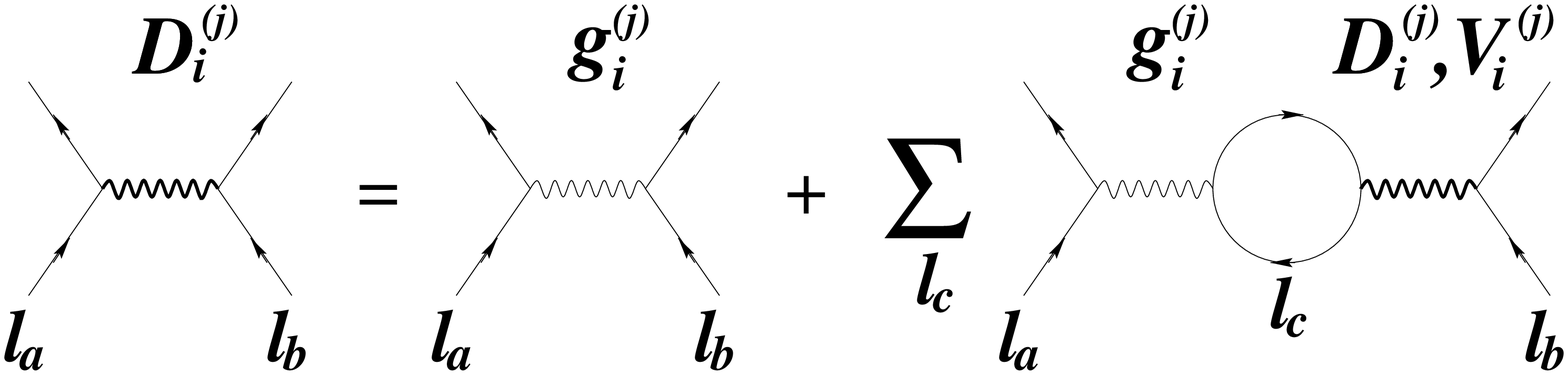}}
\end{center}
\caption{ Diagrams contributing to the screening of Coulomb 
interactions between currents $l_{a}$, $l_{b}$ of well-defined 
chirality and spin inside a rope. The second term takes into 
account the polarization of the $n$ metallic nanotubes in the rope.}
\label{rpa}
\end{figure}

Dealing for the time being with such small momentum-transfer
interactions, the dressed vertices
are given by self-consistent equations of the type shown in
Fig. \ref{rpa}.
It has to be realized that, in the case of a rope, there are
interaction vertices $D_i^{(j)}$ for currents within the same
nanotube as well as interaction vertices $V_i^{(j)}$ between
currents at different metallic nanotubes. If the number of them
is $n$, this is also the number of different terms in the sum
that appears in Fig. \ref{rpa}.

The self-consistent equations for the intratube $D_i^{(j)}$ and
the intertube $V_i^{(j)}$ vertices can be solved in the same
fashion as for the simpler case of the Luttinger model\cite{sol}.
In order to obtain a solution in the form of coupling constants, we
have taken the static approximation in the polarization
operators. We have also chosen an average
strength for the Coulomb potential $v = (e^2/2 \pi )
\log |k_c / k_0|$, where $k_c$ is the short-distance cutoff and
$k_0 \sim 10^{-3} k_c$ accounts for the long-distance average.

The appropriate bare values of the couplings are
$g_4^{(4)} = v + g$, $g_4^{(2)} = v \mp g$, $g_2^{(2)} = v + g$
and $g_2^{(4)} = v \mp g$, where $g < 0$ is the 
contribution from phonon exchange. The upper sign
applies to the armchair nanotubes and the lower sign to the
zigzag nanotubes. In the former case, the solution of the
self-consistent equations is
\begin{eqnarray}
D_4^{(4)} + D_4^{(2)} = D_2^{(2)} + D_2^{(4)} 
 &  =  &  \frac{2v}{1 + 4n v/\varepsilon_0 \pi v_F}  \label{coul} \\
D_4^{(4)} - D_4^{(2)} =  D_2^{(2)} - D_2^{(4)}   
 &  =  &  \frac{2g}{1 + 4g/\varepsilon_0 \pi v_F}   \label{phonon}
 \end{eqnarray}
In the case of the zigzag nanotubes, we have 
\begin{equation}
D_i^{(j)} = \frac{v+g+4g(nv+g)/\varepsilon_0 \pi v_F}
  {(1 + 4g/\varepsilon_0 \pi v_F)(1+4(nv+g)/\varepsilon_0 \pi v_F)}
\label{equal}
\end{equation}
for all of the four dressed couplings.
The effect of the Coulomb screening is well captured 
in the expression of the dressed vertices.
The RPA-like diagrams considered screen repulsive interactions
and enhance attractive interactions, and the latter always 
prevail for sufficiently large values of $n$. 
For $ 4|g|/\varepsilon_0 \pi v_F $ close to 1, strong non-perturbative
effects arise from the same kind of singularity already observed
in the bosonization picture\cite{lm,th2}.      
The limit $n \rightarrow 0$,  $g \rightarrow 0$ recovers the bare
vertices used in precedent studies of 
SC in isolated nanotubes \cite{cara,berk}.

{\bf Intratube scaling equations.}--- The preceding physical 
description has to be completed by incorporating
the effect of the backscattering and Umklapp effective
interactions arising from phonon exchange.
We set the signs of their respective initial strengths
(equal to $|g|$ in absolute value)
according to the above discussion, 
and take into account their  nontrivial
scaling behavior when the frequency probe is
progressively reduced. The couplings
scale according to anomalous dimensions which can be written in
terms of the dressed couplings $D_i^{(j)}$. 
To second order in the backscattering
and Umklapp couplings, the scaling equations read
\begin{eqnarray}
\partial g_1^{(2)} /\partial l  & = &
   \frac{1}{\pi v_F} ( D_4^{(2)} - D_2^{(2)} ) g_1^{(2)}
     +   \frac{1}{\pi v_F}  (  g_4^{(3)} g_1^{(3)}  \nonumber  \\
     & &   - g_2^{(1)} g_1^{(1)}  )   \label{first}     \\
\partial g_1^{(3)} /\partial l  & = &
      \frac{1}{\pi v_F} ( D_4^{(2)} + D_2^{(2)} )  g_1^{(3)}
   + \frac{1}{\pi v_F}   (  - 2 g_1^{(3)} g_1^{(1)}   \nonumber \\
  &  &   +  g_2^{(3)} g_1^{(1)}  + g_4^{(3)} g_1^{(2)}  )     \\
\partial g_2^{(1)} /\partial l  & = &
    \frac{1}{\pi v_F} ( D_4^{(2)} - D_2^{(2)} )  g_2^{(1)}
   + \frac{1}{\pi v_F}  (  g_4^{(1)} g_1^{(2)}    \nonumber     \\
   &  &    - 2 g_4^{(1)} g_2^{(1)} + g_4^{(3)} g_1^{(3)} 
   - g_4^{(3)} g_2^{(3)}  - g_1^{(2)} g_1^{(1)}  )    \\
\partial g_2^{(3)} /\partial l  & = &
    \frac{1}{\pi v_F} ( D_4^{(2)} + D_2^{(2)} ) g_2^{(3)}
  + \frac{1}{\pi v_F}  (  g_4^{(1)} g_1^{(3)}   \nonumber     \\
   &  &  - 2 g_4^{(1)} g_2^{(3)} + g_4^{(3)} g_1^{(2)} 
                   - g_4^{(3)} g_2^{(1)}  )    \\
\partial g_4^{(3)} /\partial l  & = &
   \frac{1}{\pi v_F}  2 D_4^{(2)}  g_4^{(3)}  
   +  \frac{1}{\pi v_F}   ( - g_4^{(3)} g_4^{(1)}   \nonumber   \\
    &  &   - 2 g_2^{(3)} g_2^{(1)} + g_1^{(3)} g_2^{(1)}   
   + g_2^{(3)} g_1^{(2)} + g_1^{(3)} g_1^{(2)}   ) \label{last}
\end{eqnarray}
where $l$ stands for (minus) the logarithm of the energy scale 
measured in units of the high-energy cutoff $E_c $ 
(of the order of $\sim 0.1 \; {\rm eV}$).
These equations are similar to those obtained in Ref. 
\onlinecite{berk}, except for the 
replacement of $g_2^{(2)}, g_2^{(4)}, g_4^{(4)}$
and $g_4^{(2)}$ by the respective dressed couplings. 
We apply this replacement to the rest of the equations in
the model, including those for the small momentum-transfer couplings, 
which we do not write since they are similar to those in Ref. 
\onlinecite{berk}. In this way we incorporate the {\em finite} 
renormalizations arising from the interaction among the metallic
nanotubes in the rope, which set
the initial values of the dressed couplings in accordance to Eqs. 
(\ref{coul})--(\ref{equal}).

{\bf Intertube coherence.}--- The ropes are made of nanotubes 
with different helicities and
diameters, and this compositional disorder frustrates the
formation of any phase with spin-density-wave or
charge-density-wave order in the rope. On the other hand, the
nanotubes may have strong superconducting correlations. Their
strength is given by the combination of the couplings
$D_2^{(2)}, g_1^{(1)}, g_2^{(1)}$ and $g_1^{(2)}$,
which govern the propagation of the Cooper
pairs along each nanotube. The interactions corresponding to
$g_1^{(1)}$ and $g_2^{(1)}$ are always attractive, pointing
at the enhancement of the superconducting correlations with
$s$-wave symmetry at low energies.

In the compositionally disordered ropes, the coherence is
established by the tunneling of Cooper pairs between neighboring
metallic nanotubes. The amplitude $J$ for this process has a
dependence on the energy scale characterized by the anomalous
dimension $\Delta $, given by the combination
$D_2^{(2)} + g_1^{(1)} + g_2^{(1)} + g_1^{(2)}$. The scaling
equation for $J$ is\cite{yako}
\begin{equation}
\partial J / \partial l = \Delta J + c \; t_{T}^2
\end{equation}
where $t_T$ is the intertube hopping for each particle of the
Cooper pair, of the order of $\sim 0.01 \; {\rm eV}$. We recall
that the tunneling amplitude for single electrons is in general
much smaller than this quantity, due to the mismatch 
between the Fermi points of neighboring nanotubes\cite{mkm}. This
introduces a factor of suppression that sets the
single-particle intertube hopping about three orders of
magnitude below the estimate for $t_T$ \cite{avour}.

If the generalized backscattering and Umklapp  processes 
are fixed to zero, the system is at the fixed point 
described in the bosonization picture \cite{th2}. 
That fixed point is in general unstable against chirality-breaking 
perturbations by which the system flows 
to a strong-coupling fixed point. Nevertheless, the crossover behavior is  
very slow, being characterized by a relatively small crossover 
frequency  $\omega_{xo}$. 
In a large region of the phase diagram the flow is cut off by the onset of 
the 3D coherence before $\omega_{xo}$ has been reached.  
In these conditions, the SC phase remains in the region of influence of
the weak-coupling fixed point, which governs then the main physical 
properties.
 
{\bf Results and discussion.}--- Following the above approach, 
we characterize the superconducting transition by looking for a
pole $\omega_c $ in the 3D propagator ${\cal R} (\omega )$ of 
the Cooper pairs along 
the rope. This is related to the SC response function 
${\cal R}^{(0)} (\omega )$ in the individual nanotubes by the 
Schwinger-Dyson equation $1/{\cal R} (\omega ) = 
1/{\cal R}^{(0)} (\omega ) - \tilde{J}$, where $\tilde{J}$ is the
(dimensionless) pair hopping measured in units of the high-energy 
cutoff. 

\begin{figure}
\begin{center}
\mbox{\epsfxsize 4.5cm \epsfbox{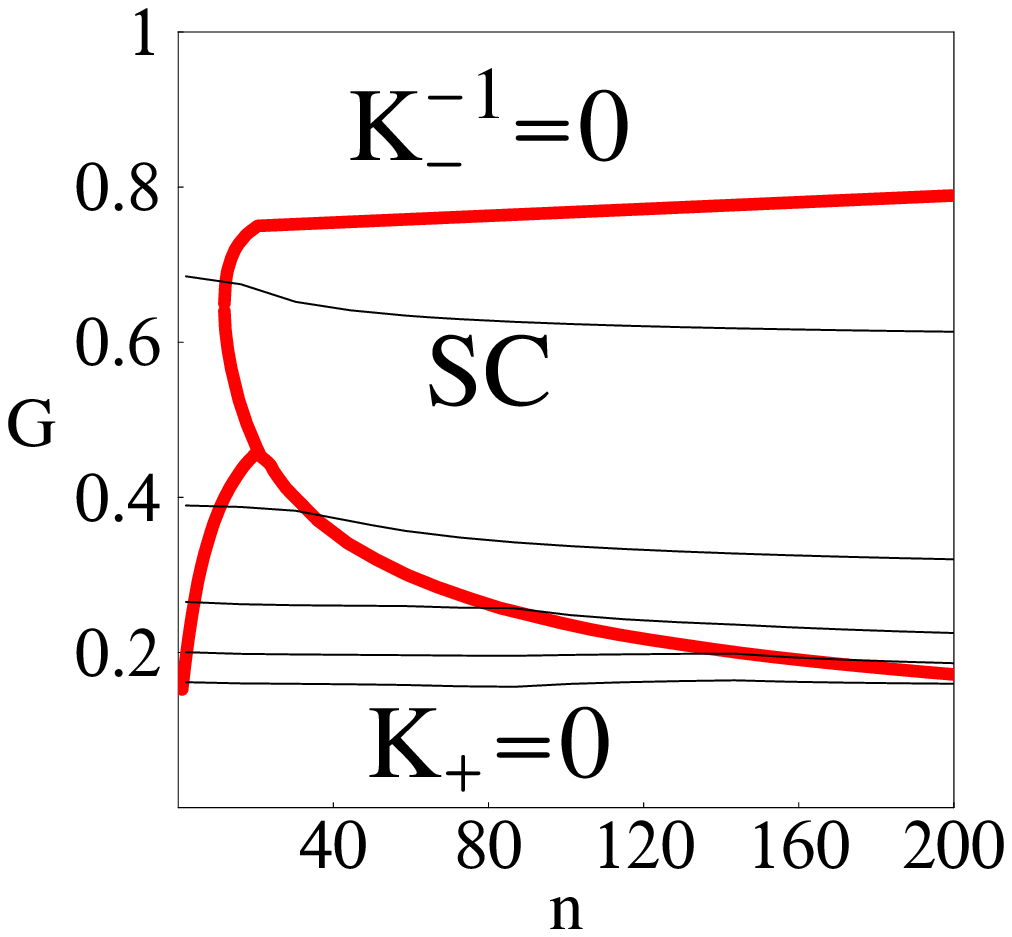}
  \epsfxsize 4.5cm \epsfbox{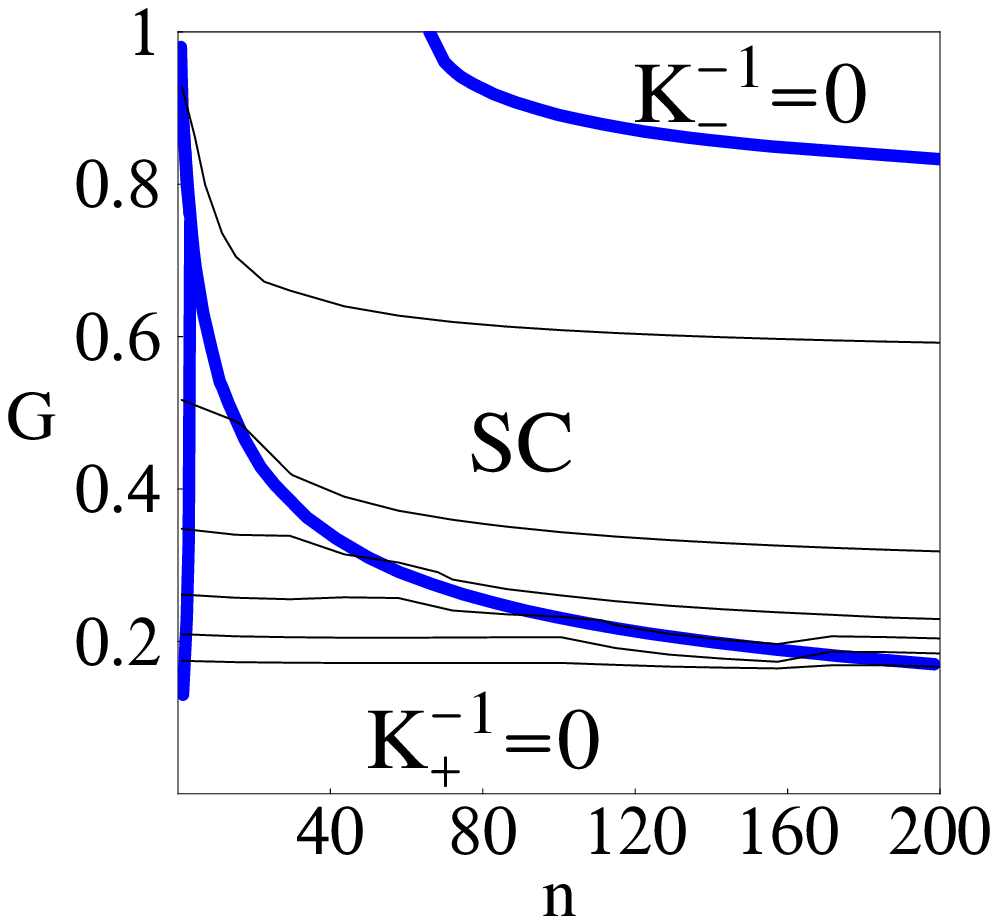}}
\end{center}
\caption{
Phase diagram of undoped armchair (left) and doped zigzag
(right) nanotubes as a function of the number of metallic
nanotubes $n$ in the rope and the strength of the
attractive interaction $G = 4|g|/\pi v_F $. Thick lines represent
the phase boundaries between a 3D phase-coherent superconducting
phase (SC) and the 1D phases characterized by the
breaking of the parameters $K_{+}$ and  $K_{-}$ (the narrow phase
to the left of the second diagram corresponds to the divergence of
the latter). Thin lines are contours of constant
critical frequency $\omega_c $, starting from above with
$\ln (E_c / \omega_c ) = 2, 4, 6, \ldots $  }
\label{PHD}
\end{figure}

Moreover, we can 
study the competition between the SC and other charge
instabilities in the rope by rebosonizing the system at each 
step in the integration of Eqs. (\ref{first})-(\ref{last}), 
considering backscattering and Umklapp couplings as small 
perturbations. The generalized Luttinger liquid parameters of the 
{\em individual}
nanotubes take the form
\begin{equation}
K_{\pm} = \sqrt{ \frac{ \pi v_F + (D_4^{(4)} \pm D_2^{(4)})
                 -  (D_2^{(2)} \pm D_4^{(2)}) }
                      { \pi v_F + (D_4^{(4)} \pm D_2^{(4)})
                 + (D_2^{(2)} \pm D_4^{(2)}) }  }   
\label{para}
\end{equation}
We obtain complementary information about the physical properties 
of the system by studying the regions where the $K_{\pm}$ parameters 
either vanish or diverge. This happens when some of the functions given 
by Eq. (\ref{para}) develops a branch-cut at some energy scale, what 
has to be understood as the onset of a phase transition in the 
system. The different phases which arise in this approach have 
been represented in Fig. \ref{PHD}, where contour lines of constant
critical scale $\omega_c $ have been also plotted.

The weak-coupling character of the SC phase allows to
estimate the transition temperatures
from the values of the pole $\omega_c $ shown in Fig. \ref{PHD}.
The comparison with the experiments reported in Ref.
\onlinecite{sup} can be made by recalling that the ropes used
there have about 100 metallic nanotubes, while the transition
temperatures measured are always below 1 K. When compared to the
high-energy cutoff, that value corresponds to $l_c \approx 7$.
From the results
shown in Fig. \ref{PHD}, we find that the strength of the
attractive interaction $ 4|g|/\pi v_{F} $
has to be $\approx 0.3$.

Given the above estimate of $g$, we conclude that the superconducting 
correlations have to disappear below a number of metallic
nanotubes $n \lesssim 50$, in ropes of the type measured in Ref.
\onlinecite{sup}. That is, for nanotubes with a typical diameter 
of $\approx 1.4 \; {\rm nm}$, 
the effect of the Coulomb interaction overcomes any 
source of effective attraction for that range of $n$. This is
in agreement with the experimental measures carried out 
by the collaboration of Ref. \onlinecite{sup} in thin ropes with
about 40 nanotubes\cite{priv}.

The lower boundary of the SC phase in Fig. \ref{PHD} almost coincides 
with the boundary at which the pair-hopping amplitude ceases to grow 
large at low energies. When this happens, the nanotubes in the rope 
start to behave as uncoupled 1D systems, what explains the disappearance 
of the superconducting transition. 
A phase with vanishing compressibility appears for 
low values of $|g|$ in the undoped system. The fact that the 
regime with $K_{+} = 0$ is only present at half-filling makes clear 
that it corresponds to the extension of the Mott insulating phase
found in the studies of carbon nanotubes with respulsive 
interactions\cite{kane,yo}. In the case of doped nanotubes, the
repulsive Umklapp interactions are not present in the system, and
the decoupling of the nanotubes in the rope leads instead to a phase
with divergent compressibility in the channel of the total charge,
as shown in Fig. \ref{PHD}.

Another phase with divergent 
compressibility arises at strong coupling in the channel 
of the mismatch of charge in the two gapless subbands.
These divergences have the same character 
of the Wentzel-Bardeen singularity, which is the natural 
counterpart of the superconducting instability\cite{lm,dme}. 
A divergent compressibility implies that the density-density 
correlations become increasingly large, pointing
at the onset of a regime where
there is no homogeneous descripition of the system.
For sufficiently strong attraction, the electronic 
charge finds then more favorable its macroscopic 
segregation than the formation of Cooper pairs. We have shown
that such a phase may be observable in ropes, either as a consequence
of a strong attractive interaction or, in doped systems,
from the decoupled behavior of the metallic nanotubes
at weak coupling, for temperatures slightly below 1 K.

\end{document}